# Anomalous Advection-Diffusion Models for Avascular Tumour Growth

Sounak Sadhukhan[1], S. K. Basu[2]

[1,2]Department of Computer Science, Banaras Hindu University, Varanasi 221005, India

**Abstract**

In this study, we model avascular tumour growth in epithelial tissue. This can help us to get a macroscopic view of the interaction between the tumour with its surrounding microenvironment and the physical changes within the tumour spheroid. This understanding is likely to assist in the development of better diagnostics, improved therapies and prognostics. In biological systems, most of the diffusive and convective processes are through cellular membranes which are porous in nature. Due to its porous nature, diffusive processes in biological systems are heterogeneous.

Fractional advection-diffusion equations are well suited to model heterogeneous biological systems; though most of the early studies did not use this fact. They modelled tumour growth with simple advection-diffusion equation or diffusion equation. We have developed two spherical models based on fractional advection-diffusion equations: one of fixed order and the other of variable order for avascular tumour. These two models are investigated from phenomenological view by measuring some parameters for characterizing avascular tumour growth over time. It is found that both the models offer realistic and insightful information for tumour growth at the macroscopic level, and approximate well the physical phenomena. The fixed-order model always overestimates clinical data like tumour radius, and tumour volume. The cell counts in both the models lie in the clinically established range. As the simulation parameters get modified due to different biochemical and biophysical processes, the robustness of the model is determined. It is found that, the sensitivity of the fixed-order model is low while the variable-order model is moderately sensitive to the parameters.

## 1. Introduction

Tumour is a lump of cells with genetic or epigenetic defects growing uncontrollably and uncoordinated manner. Tumour can be categorized into malignant and benign. Malignant tumours have the ability to invade surrounding tissues. This quality makes it more dangerous than the benign one. After the formation of a lump of cells, tumour cells depend upon the surrounding microenvironment for nutrients and oxygen as it does not have direct vascular support. In the avascular stage, the growth of a tumour is restricted due to limited supply of nutrients and oxygen. To overcome this situation a tumour needs direct blood vessel support for supply of sufficient nutrients and oxygen. Hence, it must undergo angiogenesis (*vascularization*) (Folkman (1974); Folkman (1976); Muthukkaruppan et al., 1982). During angiogenesis, the tumour induces blood vessels from proximally existing blood vessels. The new capillaries grow from the existing blood vessels towards the tumour and finally form a local vasculature for supplying oxygen and nutrients. The nutrients and oxygen supply help the tumour to grow unboundedly, which becomes fatal to the host over the time. The unbounded growth may disturb the functionalities of a critical organ. Tumour cells may detach from its source singly or collectively and enter into blood vessels (*intravasation*) (Wirtz (2009)). The circulating tumour cells in the blood vessels which succeed to survive may get arrested to the vascular endothelium to form colonies of cells (Wirtz (2009)) which eventually become life threatening to the host.

A lot of biochemical processes and biophysical stresses work coordinated manner in this phase. The tumour takes necessary nutrients and oxygen from nearby blood vessels through the extracellular matrix (ECM). As a result, a fluid flux is generated within the tumour towards its centre due to the movement of nutrients and oxygen molecules from the surrounding microenvironment. The tumour cells absorb these molecules to persist its growth. Hence, a cell flux has been formed in the outward direction throughout the tumour to expand its volume (Fig.

1(a)). The requirements of nutrients and oxygen have gradually increased with the volume of the tumour but, it can absorb necessary substances proportional to its surface area (Orme and Chaplain (1996)).

Due to over-metabolism for rapid growth, nutrients concentration, and partial pressure of oxygen gradually decrease within the innermost area of the tumour cells (Carmeliet and Jain (2000)). The deficiency level increases with the distance from the outer surface of the tumour towards its centre. At the centre, the deficiency level will be the maximum (below the threshold level). So, the tumour has an upper bound in size (critical size) up to which it can grow before it experiences nutrients and oxygen deficiency (*hypoxia*). This situation creates a layer of proliferative cells at the outer surface and a core of hypoxic cells at the centre of the tumour (Fig. 1(b)). If the deficiency is further increased, some portions of the hypoxic cells near the tumour centre become necrotic and make a core of dead cells. On the other hand, hypoxic cells release some chemical substrates (tumour angiogenic factors, growth inhibitory factors, etc.) to reduce the tumour volume (Macklin (2010)) which is absorbed by the proliferative cells. This creates another fluid flux to the outward direction. These opposite fluxes decrease the physical stress on the necrotic core and help it to expand further.

These synchronized processes divide the tumour into zones of different cells: the outer layer mostly consists of proliferative cells; the innermost layer contains only the necrotic dead cells. Another layer exists in between them named the quiescent cell layer (Fig. 1(c)). The boundaries among these layers are not clearly distinguishable (Fig. 2) (Hystad and Rofstad (1994)). With the increase in volume of the tumour, the volume of the necrotic core also increases. Once the tumour reaches its critical size, the cell flux and the fluid flux balance each other and the avascular spheroid reaches a steady state. It cannot grow further without absorption of nutrients and oxygen directly from the blood vessel (*vascularization*). The cells belonging to the quiescent layer are alive but do not divide; however, they become proliferative again if the surrounding microenvironment is suitable for them (Freyer and Schor (1987)). In this phase, an avascular tumour spheroid grows up to 2 mm in diameter before tumour angiogenesis (Zetter (1998); Carmeliet and Jain (2000); Lodish et al., 2000; Hillen and Griffioen (2007)) and it consists of approximately $10^6 - 10^7$ cells (Lodish et al., 2000; Anderson (2005)).

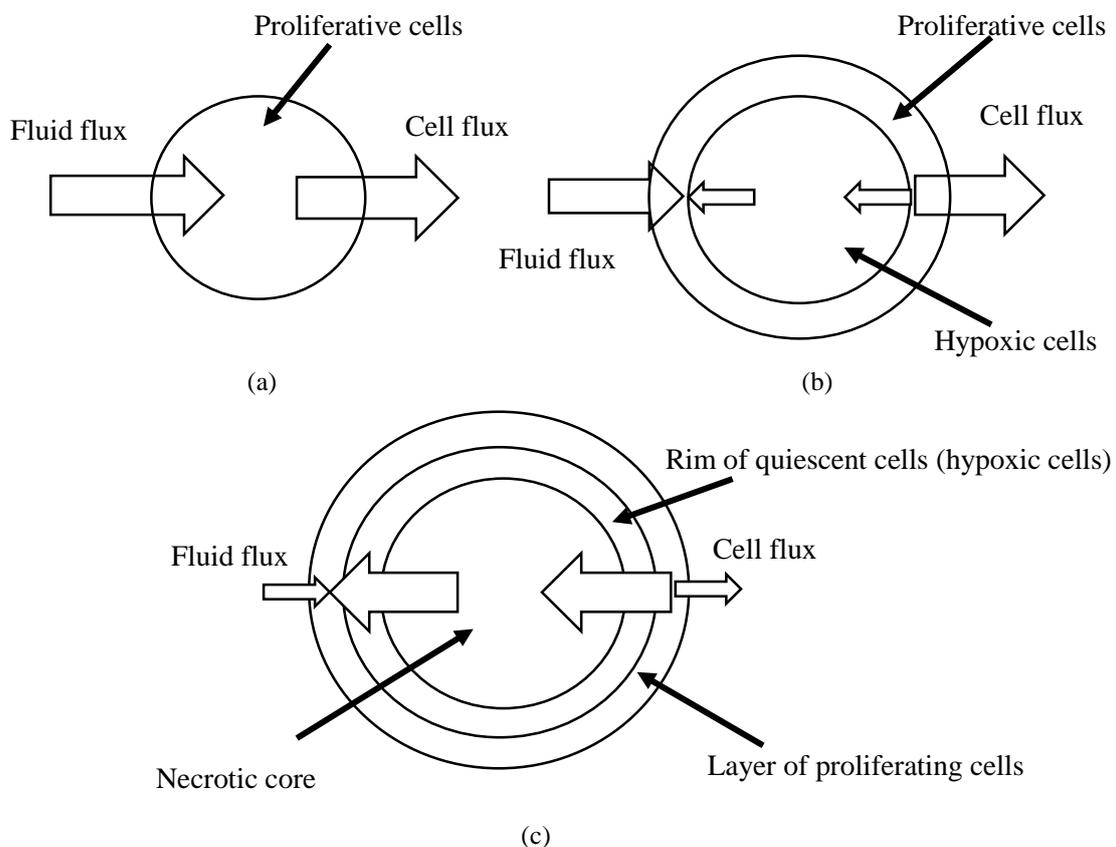

**Fig. 1** Cell flux and fluid flux in different phases ((a) early, (b) intermediate, and (c) long-time) of avascular tumour growth (adapted from Macklin (2010)).

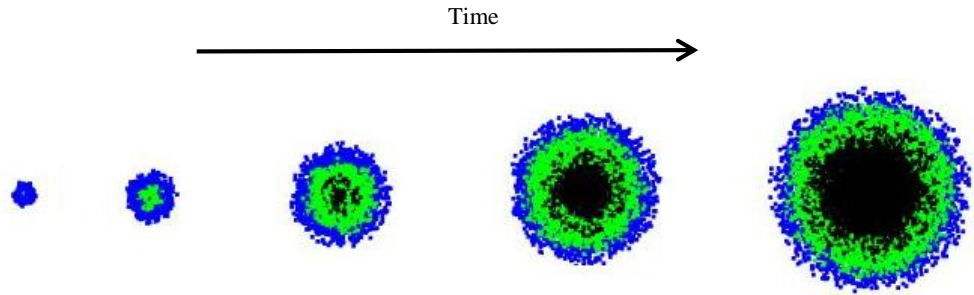

**Fig. 2** 2-D view of avascular tumour at different time intervals (black, green, and blue coloured cells are indicating necrotic, quiescent, and proliferative cells respectively).

Avascular tumours have great clinical importance. Tumour volume and volume doubling time (VDT) are the important parameters from clinical aspects. VDT refers to the time a tumour takes to double its volume. Shorter VDT refers to the aggressiveness of the tumour (Usuda et al., 1994). VDT may vary between different types of tumours. It may also vary for the same type of tumours in two different human beings. As an example, for the lung tumours, VDT may vary between 150 to 400 days, whereas for small cell lung tumour VDT is less than 100 days (Kanashiki et al., 2012). But there are exceptions too; that is, the tumour spheroid which remains stable for last two years does not guarantee that it is a benign tumour as some of the malignant tumours have a long VDT (Kanashiki et al., 2012). It is clinically said that the cancer patients whose VDT are over 200 days have a better prognosis. Oncologists have estimated that, almost 30 VDTs are required for a tumour to grow up to 2 mm in diameter (Oncologist's Note: https://notesofoncologist.com/2018/02/26/how-fast-do-tumours-grow/).

Tumour growth modelling over the last decades has become a growing field of research. Intricacies of most of the bio-chemical and bio-physical phenomena regarding tumour growth and its progressions are less understood. Researchers have developed various models to explain avascular tumour growth from different perspectives. Greenspan (1972) have developed the first-ever tumour growth model with proliferative, quiescent, and necrotic cell zones that were regulated by mitotic inhibitors and necrotic decomposition. The model incorporated various phases of growth leading to saturation. Later, many researchers have adapted this model framework and tried different modifications and advancements on it.

Ward and King (1997) developed a model for avascular tumour growth that is based on diffusion theory. In this model the tumour growth is directly influenced by the nutrient concentrations in the microenvironment. The model considered that the cell death is a gradual process and the death rate does not depend on the local nutrients level. The tumour volume changes with the cell growth and death which creates a velocity field in the continuum of live cells. In another study, they (Ward and King (1999)) extended the previous work (Ward and King (1997)) by adding a physical mechanism for growth saturation. In this work, authors have assumed that, for reproduction, a tumour cell depends on cellular materials such as proteins, DNA, and lipid (due to the breakdown of necrosis or external tissue *in vivo*), and nutrients and other growth factors (*in vitro*). Two mechanisms have been proposed for the reduction of cellular materials, and loss of volumes: first includes the leakage of cellular material by diffusion to the external matrix (ECM) and second, it includes the consumption of cellular materials for constructing new cells. Sherratt and Chaplain (2001) developed a 1-D spatio-temporal model for avascular tumour which grows in epithelium. The model was developed in terms of different cell densities like, proliferative, quiescent, and necrotic with contact inhibition. The tumour growth is driven by the cell movements due to nutrient concentrations. The model considered that the different cell-layer boundaries are not sharply distinguishable.

Smallbone et al., (2005), Kiran et al., (2009), Grimes et al., (2014), and Grimes et al., (2016) considered that different cell layers in the tumour are distinctly identifiable. Smallbone et al., (2005) developed a three-dimensional model to examine the effect of acidosis on avascular tumour cell populations. Kiran et al., (2009) established a continuum model based on diffusion. They considered that the tumour is spherical in shape and its growth depends on the diffusive nutrients across the contiguous tissue. Tumour cells die due to apoptosis as well as necrosis. The authors also measured the radii of quiescent and necrotic zones and validated the measurement with the available *in vitro* tumour growth data and Gompertzian empirical relationship parameters. Grimes et al., (2014) built reaction-diffusion models to study the effect of oxygen consumption rates in tumour cells at different oxygen tensions. In another study, Grimes et al., (2016) examined the oxygen dynamics within the tumour spheroids and found that the cellular doubling time and oxygen consumption rate can be enough to define a tumour growth.

In this study, we focus on avascular tumour in epithelial tissue and consider that it is spherical in shape. Nutrients and oxygen in the microenvironment synthesize the structural support of the tumour cells and stimulates the tumour cells to proliferate and migrate. The cell migration takes place due to both diffusive as well as convective processes. Diffusion processes in biological systems are anomalous (Gal and Weihs (2010)) as the transportations in a biological system take place through the cellular membranes which are porous in nature. The fractional advection-diffusion equation (FADE) (Benson et al., (2000a); Benson et al., (2000b)) is one of the popular techniques to model anomalous transportation. In the first phase of the study, we develop a spherical model based on simple ADE for avascular tumour growth assuming overlapped boundaries between different cell layers. Then the model is modified by including fixed order-FADE (FO-FADE). The FO-FADE model is simulated to examine its behaviour with a set of parameters and suitable initial and boundary conditions.

Biological systems are very complex in nature, the structure of the cellular membrane and also the external fluid fields change over time. So, de Azevedo et al., (2006) and Atangana and Secer (2013) suggested use of variable-order FADE (VO-FADE) in this context. In the second phase of the study, modifications of FO-FADE model are made by including VO-FADE. A suitable parameter is also included to capture skewed diffusion in both the models. We study both of these models (FO-FADE and VO-FADE) from phenomenological and clinical point of views by measuring some parameters (tumour radius, tumour volume, total number of tumour cells) over time. We also calculate tumour VDT for both the models by varying proliferation rate.

This paper is organized as follows: section 2 describes a simple ADE based spherical model; section 3 describes modification of the simple ADE model based on anomalous diffusion. Section 3.1 describes the FO-FADE model; and section 3.2 discusses the model based on VO-FADE. Different characteristics, clinical effectiveness and robustness of these two models are discussed in section 3.3. Section 4 concludes the paper.

## 2. Simple Advection-Diffusion Model

In this study, our focus is on the avascular tumour growth in the epithelium cells. In this section, we develop a 3-D model in terms of three different cell densities. We use $p(x, y, z, t)$, $q(x, y, z, t)$, and $n(x, y, z, t)$ to denote the concentration of proliferative, quiescent, and necrotic cells respectively. The centre of the tumour is at $x = 0$, $y = 0$, and $z = 0$. The simple ADE model is inspired by the study of Sherratt and Chaplain (2001). The tumour growth obeys Fick's law involving convective and diffusive cell migrations. The convection is due to the movement of the extracellular matrices (ECM) surrounding the tumour. $v_e$ is the convective velocity of ECM. Necrotic cells cannot migrate but the proliferative and quiescent cells have migration capabilities; the interplay among them is triggered due to the changes in nutrient concentrations and partial oxygen pressure. The model considered that boundaries of different cell layers are not distinguishable. Therefore, the cell fluxes corresponding to the proliferative and quiescent concentrations as $[D_p (\partial^2 p/\partial x^2 + \partial^2 p/\partial y^2 + \partial^2 p/\partial z^2) - v_e (\partial p/\partial x + \partial p/\partial y + \partial p/\partial z)]$ and $[D_q (\partial^2 q/\partial x^2 + \partial^2 q/\partial y^2 + \partial^2 q/\partial z^2) - v_e (\partial q/\partial x + \partial q/\partial y + \partial q/\partial z)]$ respectively. When the nutrients and the partial pressure of oxygen reduces below a certain level within the tumour, hypoxic cells near the centre become necrotic. For necrotic cells, there is no flux as they are collection of dead cells. We also assume that proliferative and quiescent cells have the same rate of apoptosis *i.e.*, $d_p = d_q$.

In this model, we consider only two types of independent variables, nutrient concentrations ($c_n(x, y, z, t)$) and oxygen partial pressure ($P_o(x, y, z, t)$). The model ignores the fact that hypoxic cells releases a number of chemical substrates to reduce tumour volume. Transformation rates of proliferative cells into the quiescent and quiescent cells into necrotic are influenced by the nutrient concentrations ($c_n$) and partial pressure of oxygen ($P_o$). As the tumour is assumed spherical in shape, nutrients and oxygen move through all the cells from the outer surface towards the centre of the tumour. Nutrients and oxygen during diffusion from the source (blood vessels) have to pass through ECM. Hence, we consider the fluxes corresponding to the nutrients and partial pressure of oxygen as $[D_n (\partial^2 c_n/\partial x^2 + \partial^2 c_n/\partial y^2 + \partial^2 c_n/\partial z^2) - v_e (\partial c_n/\partial x + \partial c_n/\partial y + \partial c_n/\partial z)]$ and $[D_o (\partial^2 P_o/\partial x^2 + \partial^2 P_o/\partial y^2 + \partial^2 P_o/\partial z^2) - v_e (\partial P_o/\partial x + \partial P_o/\partial y + \partial P_o/\partial z)]$, respectively. Proliferative and quiescent cells are both alive, but only the proliferative cells divide. Hence, the model considers that the proliferative cells consume more nutrients and oxygen than the quiescent cells. The model is represented by the PDEs (1).

$$\begin{aligned}
\frac{\partial p}{\partial t} &= \underbrace{\left(D_p \nabla^2 p - v_e \nabla p\right)}_{\text{net diffusion flux}} + \underbrace{\alpha h_1(c_n, P_o)p}_{\text{source term}} - \underbrace{\beta h_2(c_n, P_o)p - d_p p}_{\text{decay term}} \\
\frac{\partial q}{\partial t} &= \left(D_q \nabla^2 q - v_e \nabla q\right) + \overbrace{\beta h_2(c_n, P_o)p}^{\text{source of necrotic cells}} - \gamma h_3(c_n, P_o)q - d_q q \\
\frac{\partial n}{\partial t} &= \gamma h_3(c_n, P_o)q \\
\frac{\partial c_n}{\partial t} &= \underbrace{\left(D_n \nabla^2 c_n - v_e \nabla c_n\right)}_{\text{net diffusion flux}} + \underbrace{\mu_n c_n}_{\text{source term}} - \underbrace{w_1 c_n - w_2 p c_n - w_3 q c_n}_{\text{decay term}} \\
\frac{\partial P_o}{\partial t} &= \left(D_o \nabla^2 P_o - v_e \nabla P_o\right) + \mu_o P_o - k_1 P_o - k_2 p P_o - k_3 q P_o
\end{aligned} \quad (1)$$

In the above system of equations (1), $D_p$, $D_q$, $D_n$ and $D_o$ are the diffusion coefficients of proliferative cells, quiescent cells, nutrients and oxygen respectively; $\alpha$, $\beta$, and $\gamma$ are the scalars controlling the production and loss terms of proliferative, quiescent, and necrotic cells respectively. We also denote $\mu_n$ and $\mu_o$ as the production rates of nutrients and oxygen respectively in the microenvironment. $w_1$, $w_2$, $w_3$ and $k_1$, $k_2$, $k_3$ describe the losses of nutrients and oxygen respectively.

We need to define three functions $h_1$, $h_2$, and $h_3$. $h_1$, $h_2$, and $h_3$ are associated with the source term of proliferative, quiescent, and necrotic cell concentrations. $h_1$, $h_2$, and $h_3$ vary with nutrients and oxygen concentration as represented in Equation (2).

$$h_1(c_n, P_o) = 1 + 0.1(c_n P_o); \quad h_2(c_n, P_o) = \frac{e^{-\frac{P_o}{\varepsilon_1}}}{1 + e^{-\frac{c_n}{\varepsilon_1}}}; \quad h_3(c_n, P_o) = \frac{e^{-\frac{P_o}{\varepsilon_2}}}{1 + e^{-\frac{c_n}{\varepsilon_2}}}; \quad \text{where } 0 < \varepsilon_1, \varepsilon_2 < 1. \quad (2)$$

Using spherical symmetry, Equations (1a) through (1e) are reduced to Equation (3). Hence, proliferative cells, quiescent cells, necrotic cells, nutrients concentration, and oxygen partial pressure are represented by $p(r, t)$, $q(r, t)$, $n(r, t)$, $c_n(r, t)$, and $P_o(r, t)$ respectively.

$$\begin{aligned}
\frac{\partial p}{\partial t} &= \frac{\partial}{\partial r}\left(D_p \frac{\partial p}{\partial r} - v_e p\right) + \frac{2}{r}\left(D_p \frac{\partial p}{\partial r} - v_e p\right) + \alpha h_1(c_n, P_o)p - \beta h_2(c_n, P_o)p - d_p p \\
\frac{\partial q}{\partial t} &= \frac{\partial}{\partial r}\left(D_q \frac{\partial q}{\partial r} - v_e q\right) + \frac{2}{r}\left(D_q \frac{\partial q}{\partial r} - v_e q\right) + \beta h_2(c_n, P_o)p - \gamma h_3(c_n, P_o)q - d_q q \\
\frac{\partial n}{\partial t} &= \gamma h_3(c_n, P_o)q \\
\frac{\partial c_n}{\partial t} &= \frac{\partial}{\partial r}\left(D_n \frac{\partial c_n}{\partial r} - v_e c_n\right) + \frac{2}{r}\left(D_n \frac{\partial c_n}{\partial r} - v_e c_n\right) + \mu_n c_n - w_1 c_n - w_2 p c_n - w_3 q c_n \\
\frac{\partial P_o}{\partial t} &= \frac{\partial}{\partial r}\left(D_o \frac{\partial P_o}{\partial r} - v_e P_o\right) + \frac{2}{r}\left(D_o \frac{\partial P_o}{\partial r} - v_e P_o\right) + \mu_o P_o - k_1 P_o - k_2 p P_o - k_3 q P_o
\end{aligned} \quad (3)$$

## 3. Anomalous Advection-Diffusion Models

Cells interact with its environment through the cell membranes which are porous in nature. It is reported that transportations through porous media are usually anomalous or non-Fickian (Zhang et al., 1994; Ellsworth et al., 1996; Pachepsky et al., 2000). For the non-Fickian diffusive processes, diffusion coefficients depend on the medium (Caputo and Cametti (2008)). Therefore, we modify the system of Equations (3) based on this fact of anomalous transportation across porous cell boundary as many researchers did not utilize this fact in their studies. FO-FADE (in section 3.1) as well as VO-FADE (in section 3.2) both are studied with this anomaly.

## 3.1 Fixed Fractional Order ADE Model

Sometimes fluids may carry solid particles during transportation and this may close some of the pores in the membrane. This phenomenon illustrates that the diffusion processes through cell membranes are medium dependent, indicating a space dependant diffusion coefficient (Caputo and Cametti (2008)). We propose a continuum model from phenomenological point of view, based on coupled FO-FADE (fractional order $\phi$ and $\xi$ where $1 < \phi \leq 2$ and $0 < \xi \leq 1$) for different types of tumour cells (proliferative, quiescent, and necrotic), nutrients concentration and oxygen partial pressure. In this case, space fractional derivatives are included. We also include a suitable parameter ($\psi$) to handle the skewness in the diffusion. We consider diffusion coefficients as functions of nutrients and oxygen partial pressure as both of these are varies with spatial and temporal domains. We study the model with a set of parameters (mentioned in Table 1) and suitable initial and boundary conditions.

As the medium of diffusion and convection is porous, it is assumed that the diffusion coefficients and the convective velocity are related as

$$(Diffusion\ coefficients) \propto (convection\ velocity)^\rho, \text{ where } 1 \leq \rho \leq 2 \tag{4}$$

According to (4), we consider,

$$\begin{aligned}
v_e(c_n, P_o) &= v_{e0}h(c_n, P_o); \\
D_p(c_n, P_o) &= D_{p0}h^\rho(c_n, P_o); \\
D_q(c_n, P_o) &= D_{q0}h^\rho(c_n, P_o); \\
D_n(c_n, P_o) &= D_{n0}h^\rho(c_n, P_o); \\
D_o(c_n, P_o) &= D_{o0}h^\rho(c_n, P_o);
\end{aligned} \tag{5}$$

FADE model is characterized by non-local transport behaviour over large distances via a convolutional fractional-derivative. Cushman and his team have claimed in their study (Cushman et al., (2000)) that FADE is a special case of classical mass balance equation. (Schumer et al., 2001) have used a generalized central limit theorem to sum up the length of particle jumps during their random walk through a heterogeneous porous medium which illustrates the behaviour of the FADE model. According to (Zhang et al., 2006a, Zhang et al., 2006b), fractional flux advection–diffusion equation in spherical domain with variable diffusion coefficients and convection velocity (Equation (5)) can be written as in Equation (6).

$$\begin{aligned}
\frac{\partial p}{\partial t} &= D_{p0}h^\rho \frac{\partial^\phi p}{\partial r^\phi} - v_{e0}h \frac{\partial^\xi p}{\partial r^\xi} + \frac{2}{r}\left(D_{p0}h^\rho \frac{\partial^\xi p}{\partial r^\xi} - v_{e0}hp\right) + D_{p0}\rho\left(\frac{\partial^\xi h}{\partial c_n^\xi}\frac{\partial^\xi c_n}{\partial r^\xi} + \frac{\partial^\xi h}{\partial P_o^\xi}\frac{\partial^\xi P_o}{\partial r^\xi}\right)\frac{\partial^\xi p}{\partial r^\xi} \\
&\quad - v_{e0}\left(\frac{\partial^\xi h}{\partial c_n^\xi}\frac{\partial^\xi c_n}{\partial r^\xi} + \frac{\partial^\xi h}{\partial P_o^\xi}\frac{\partial^\xi P_o}{\partial r^\xi}\right)p + \alpha h_1(c_n, P_o)p - \beta h_2(c_n, P_o)p - d_p p \\
\frac{\partial q}{\partial t} &= D_{q0}h^\rho \frac{\partial^\phi q}{\partial r^\phi} - v_{e0}h \frac{\partial^\xi q}{\partial r^\xi} + \frac{2}{r}\left(D_{q0}h^\rho \frac{\partial^\xi q}{\partial r^\xi} - v_{e0}hq\right) + D_{q0}\rho\left(\frac{\partial^\xi h}{\partial c_n^\xi}\frac{\partial^\xi c_n}{\partial r^\xi} + \frac{\partial^\xi h}{\partial P_o^\xi}\frac{\partial^\xi P_o}{\partial r^\xi}\right)\frac{\partial^\xi q}{\partial r^\xi} \\
&\quad - v_{e0}\left(\frac{\partial^\xi h}{\partial c_n^\xi}\frac{\partial^\xi c_n}{\partial r^\xi} + \frac{\partial^\xi h}{\partial P_o^\xi}\frac{\partial^\xi P_o}{\partial r^\xi}\right)q + \beta h_2(c_n, P_o)p - \gamma h_3(c_n, P_o)q - d_q q \\
\frac{\partial n}{\partial t} &= \gamma h_3(c_n, P_o)q \\
\frac{\partial c_n}{\partial t} &= D_{n0}h^\rho \frac{\partial^\phi c_n}{\partial r^\phi} - v_{e0}h \frac{\partial^\xi c_n}{\partial r^\xi} + \frac{2}{r}\left(D_{n0}h^\rho \frac{\partial^\xi c_n}{\partial r^\xi} - v_{e0}hc_n\right) + D_{n0}\rho\left(\frac{\partial^\xi h}{\partial c_n^\xi}\frac{\partial^\xi c_n}{\partial r^\xi} + \frac{\partial^\xi h}{\partial P_o^\xi}\frac{\partial^\xi P_o}{\partial r^\xi}\right)\frac{\partial^\xi c_n}{\partial r^\xi} \\
&\quad - v_{e0}\left(\frac{\partial^\xi h}{\partial c_n^\xi}\frac{\partial^\xi c_n}{\partial r^\xi} + \frac{\partial^\xi h}{\partial P_o^\xi}\frac{\partial^\xi P_o}{\partial r^\xi}\right)c_n + \mu_n c_n - w_1 c_n - w_2 c_n p - w_3 c_n q \\
\frac{\partial P_o}{\partial t} &= D_{o0}h^\rho \frac{\partial^\phi P_o}{\partial r^\phi} - v_{e0}h \frac{\partial^\xi P_o}{\partial r^\xi} + \frac{2}{r}\left(D_{o0}h^\rho \frac{\partial^\xi P_o}{\partial r^\xi} - v_{e0}hP_o\right) + D_{o0}\rho\left(\frac{\partial^\xi h}{\partial c_n^\xi}\frac{\partial^\xi c_n}{\partial r^\xi} + \frac{\partial^\xi h}{\partial P_o^\xi}\frac{\partial^\xi P_o}{\partial r^\xi}\right)\frac{\partial^\xi P_o}{\partial r^\xi} \\
&\quad - v_{e0}\left(\frac{\partial^\xi h}{\partial c_n^\xi}\frac{\partial^\xi c_n}{\partial r^\xi} + \frac{\partial^\xi h}{\partial P_o^\xi}\frac{\partial^\xi P_o}{\partial r^\xi}\right)P_o + \mu_o P_o - k_1 P_o - k_2 P_o p - k_3 P_o q
\end{aligned} \tag{6}$$

where, $v_{e0}$, $D_{p0}$, $D_{q0}$, $D_{n0}$, and $D_{o0}$ are the constant coefficients; $\frac{\partial^\phi p}{\partial r^\phi}$ and $\frac{\partial^\xi p}{\partial r^\xi}$ are the fractional derivatives with respect to $r$ and $1 < \phi \leq 2$ and $0 < \xi \leq 1$ ($\xi = \phi - 1$) are the fractional orders. The derivatives have been approximated by (Benson et al., 2000a; Huang et al., 2008) as shown in Equations (7 and 8).

$$\frac{\partial^\phi p}{\partial r^\phi} \approx \left(\left(\frac{1+\psi}{2}\right) \mathcal{D}_L^\phi(p) + \left(\frac{1-\psi}{2}\right) \mathcal{D}_R^\phi(p)\right), \quad 1 < \phi \leq 2 \tag{7}$$

and $\quad \frac{\partial^\xi p}{\partial r^\xi} \approx \left(\left(\frac{1+\psi}{2}\right) \mathcal{D}_L^\xi(p) - \left(\frac{1-\psi}{2}\right) \mathcal{D}_R^\xi(p)\right), \quad 0 < \xi \leq 1 \tag{8}$

$\psi$ ($-1 \leq \psi \leq 1$) is the skewness parameter which controls the bias of the diffusion. It reflects the relative weight of forward versus backward transition probability. $\phi = 2$ and $\psi = 0$ indicate no skewness in the diffusion (Fick's diffusion). If $\psi < 0$, the dispersion is skewed backward: a slow evolving contaminant plume followed by a heavy tail; whereas $\psi > 0$ shows a forward dispersion: a fast evolving contaminant plume followed by a light tail.

$\mathcal{D}_L^\phi$ ($\mathcal{D}_L^\xi$) and $\mathcal{D}_R^\phi$ ($\mathcal{D}_R^\xi$) in (7) and (8) are the left- and right-handed fractional derivatives respectively. $L = 1$ and $R = 2$ (in $\mathcal{D}_L^\phi$ and $\mathcal{D}_R^\phi$) are the corresponding lower and upper bounds of $\phi$. $L = 0$ and $R = 1$ are the corresponding lower and upper bounds of $\xi$ in $\mathcal{D}_L^\xi$ and $\mathcal{D}_R^\xi$. The left- and right-hand derivatives are as in Equations (9 and 10).

$$\mathcal{D}_L^\phi = \frac{\partial^\phi}{\partial(r)^\phi} \quad \text{and} \quad \mathcal{D}_R^\phi = \frac{\partial^\phi}{\partial(-r)^\phi} \tag{9}$$

$$\mathcal{D}_L^\xi = \frac{\partial^\xi}{\partial(r)^\xi} \quad \text{and} \quad \mathcal{D}_R^\xi = \frac{\partial^\xi}{\partial(-r)^\xi} \tag{10}$$

In Equation (6), we use $h$ instead of $h(c_n, P_o)$, and $h^\rho$ instead of $h^\rho(c_n, P_o)$. We also consider,

$$h(c_n, P_o) = exp(\theta - c_n P_o), \theta \text{ is a constant, } \theta > 0 \tag{11}$$

We assume that the minimum distance from the tumour centre ($r = 0$) to the nearest blood vessel is $l$. Hence, a spherical domain of radius $l$ is considered in which the tumour has grown. Here, $r$ is the radial direction from the centre ($r = 0$) towards the boundary ($r = l$) of the spherical shaped domain. The tumour cells and the fluid concentrations within the spherical domain from $r = 0$ to $r = l$ is under consideration. To solve this system of Equations, we first non-dimensionalize Equation (6) by rescaling the distance $l$ with time $\tau = l^2/D_{o0}$. Proliferative cell, quiescent cell, necrotic cell, nutrients concentrations, and partial pressure of oxygen are rescaled with $p_0$, $q_0$, $n_0$, $c_0$, and $P_1$ respectively (where $p_0$, $q_0$, $n_0$, $c_0$, and $P_1$ are the appropriate reference variables). Therefore,

$$p^* = \frac{p}{p_0}, \quad q^* = \frac{q}{q_0}, \quad n^* = \frac{n}{n_0}, \quad c_n^* = \frac{c_n}{c_0}, \quad P_o^* = \frac{P_o}{P_1}, \quad t^* = \frac{t}{\tau}.$$

We get the new system of Equations (12) (by dropping the stars),

$$\begin{aligned}
\frac{\partial p}{\partial t} &= \left(D_1 h^\rho \frac{\partial^\phi p}{\partial r^\phi} - vh \frac{\partial^\xi p}{\partial r^\xi}\right) + \frac{2}{r}\left(D_1 h^\rho \frac{\partial^\xi p}{\partial r^\xi} - vhp\right) + D_1 \rho \left(\frac{\partial^\xi h}{\partial c_n^\xi} \frac{\partial^\xi c_n}{\partial r^\xi} + \frac{\partial^\xi h}{\partial P_o^\xi} \frac{\partial^\xi P_o}{\partial r^\xi}\right) \frac{\partial^\xi p}{\partial r^\xi} \\
&\quad - v\left(\frac{\partial^\xi h}{\partial c_n^\xi} \frac{\partial^\xi c_n}{\partial r^\xi} + \frac{\partial^\xi h}{\partial P_o^\xi} \frac{\partial^\xi P_o}{\partial r^\xi}\right) p + \alpha h_1(c_n, P_o)p - \beta h_2(c_n, P_o)p - d_p p \\
\frac{\partial q}{\partial t} &= \left(D_2 h^\rho \frac{\partial^\phi q}{\partial r^\phi} - vh \frac{\partial^\xi q}{\partial r^\xi}\right) + \frac{2}{r}\left(D_2 h^\rho \frac{\partial^\xi q}{\partial r^\xi} - vhq\right) + D_2 \rho \left(\frac{\partial^\xi h}{\partial c_n^\xi} \frac{\partial^\xi c_n}{\partial r^\xi} + \frac{\partial^\xi h}{\partial P_o^\xi} \frac{\partial^\xi P_o}{\partial r^\xi}\right) \frac{\partial^\xi q}{\partial r^\xi} \\
&\quad - v\left(\frac{\partial^\xi h}{\partial c_n^\xi} \frac{\partial^\xi c_n}{\partial r^\xi} + \frac{\partial^\xi h}{\partial P_o^\xi} \frac{\partial^\xi P_o}{\partial r^\xi}\right) q + \eta h_2(c_n, P_o)p - \gamma h_3(c_n, P_o)q - d_q q \\
\frac{\partial n}{\partial t} &= \omega h_3(c_n, P_o)q \\
\frac{\partial c_n}{\partial t} &= \left(D_3 h^\rho \frac{\partial^\phi c_n}{\partial r^\phi} - vh \frac{\partial^\xi c_n}{\partial r^\xi}\right) + \frac{2}{r}\left(D_3 h^\rho \frac{\partial^\xi c_n}{\partial r^\xi} - vhc_n\right) + D_3 \rho \left(\frac{\partial^\xi h}{\partial c_n^\xi} \frac{\partial^\xi c_n}{\partial r^\xi} + \frac{\partial^\xi h}{\partial P_o^\xi} \frac{\partial^\xi P_o}{\partial r^\xi}\right) \frac{\partial c_n}{\partial r} \\
&\quad - v\left(\frac{\partial^\xi h}{\partial c_n^\xi} \frac{\partial^\xi c_n}{\partial r^\xi} + \frac{\partial^\xi h}{\partial P_o^\xi} \frac{\partial^\xi P_o}{\partial r^\xi}\right) c_n + \mu_n c_n - w_1 c_n - w_2 c_n p - w_3 c_n q \\
\frac{\partial P_o}{\partial t} &= \left(h^\rho \frac{\partial^\phi P_o}{\partial r^\phi} - vh \frac{\partial^\xi P_o}{\partial r^\xi}\right) + \frac{2}{r}\left(h^\rho \frac{\partial^\xi P_o}{\partial r^\xi} - vhP_o\right) + \rho \left(\frac{\partial^\xi h}{\partial c_n^\xi} \frac{\partial^\xi c_n}{\partial r^\xi} + \frac{\partial^\xi h}{\partial P_o^\xi} \frac{\partial^\xi P_o}{\partial r^\xi}\right) \frac{\partial^\xi P_o}{\partial r^\xi} \\
&\quad - v\left(\frac{\partial^\xi h}{\partial c_n^\xi} \frac{\partial^\xi c_n}{\partial r^\xi} + \frac{\partial^\xi h}{\partial P_o^\xi} \frac{\partial^\xi P_o}{\partial r^\xi}\right) P_o + \mu_o P_o - k_1 P_o - k_2 P_o p - k_3 P_o q
\end{aligned} \tag{12}$$

where, $\alpha^* = \frac{\alpha l^2}{D_{oo}}$, $\beta^* = \frac{\beta l^2}{D_{oo}}$, $d_p^* = \frac{d_p l^2}{D_{oo}}$, $\eta^* = \frac{\beta l^2}{D_{oo} q_0}$, $d_q^* = \frac{d_q l^2}{D_{oo}}$, $\gamma^* = \frac{\gamma l^2}{D_{oo}}$, $\omega^* = \frac{\gamma l^2}{D_{oo} n_0}$, $\mu_n^* = \frac{\mu_n l^2}{D_{oo}}$, $w_1^* = \frac{w_1 l^2}{D_{oo}}$, $w_2^* = \frac{w_2 l^2}{D_{oo}}$, $w_3^* = \frac{w_3 l^2}{D_{oo}}$, $\mu_o^* = \frac{\mu_o l^2}{D_{oo}}$, $k_1^* = \frac{k_1 l^2}{D_{oo}}$, $k_2^* = \frac{k_2 l^2}{D_{oo}}$, $k_3^* = \frac{k_3 l^2}{D_{oo}}$, $v^* = \frac{v_{eo} l}{D_{oo}}$, $D_1^* = \frac{D_{po}}{D_{oo}}$, $D_2^* = \frac{D_{qo}}{D_{oo}}$, $D_3^* = \frac{D_{no}}{D_{oo}}$.

To solve the system of equations, initial and boundary conditions are required. We assume that, initially tumour microenvironment is full of nutrients and oxygen and there is no trace of quiescent and necrotic cells in the domain of interest; only proliferative cells are present. With the above assumptions, the following initial conditions are found:

$$p(r, 0) = \frac{e^{-\frac{r}{10}}}{100}, q(r, 0) = 0, n(r, 0) = 0, c_n(r, 0) = 1, \text{ and } P_o(r, 0) = 1, \text{ where } r \in [0, l] \quad (13)$$

We further consider that, at any time, there is no cell flux at the boundary (at $r = 0$ and $r = l$) of the domain, but the boundaries are nutrients and oxygen rich. From these assumptions, we find the boundary conditions:

$p(0, t) = p(l, t) = 0, q(0, t) = q(l, t) = 0, c_n(0, t) = c_n(l, t) = 1,$ and $P_o(0, t) = P_o(l, t) = 1,$ where $t \in [0, T]$ (no boundary condition is needed for $n$), $T$ is the total simulation time. (14)

### 3.1.1 Estimation of Parameters

For the purpose of simulation, we consider that at the very beginning, nutrients and oxygen levels throughout the tumour domain remain uniform. That is, at every point in the spherical domain receive equal amount of nutrients and oxygen from the source (nearby blood vessels). We simulate the process with the help of Equations (12) with the initial (13) and boundary conditions (14). The values of the parameters used are described in Table 1.

Table 1: Parameter values

| Symbols | Quantity | Suggested in literature | Used values |
|---|---|---|---|
| $D_p$ | diffusivity of proliferative tumour cells | $6.90 \times 10^{-9}$ cm$^2$s$^{-1}$ to $3.50 \times 10^{-11}$ cm$^2$s$^{-1}$ (Sherratt and Murray (1991)) | $3.50 \times 10^{-11}$ cm$^2$s$^{-1}$ |
| $D_q$ | diffusivity of quiescent tumour cells | $6.90 \times 10^{-9}$ cm$^2$s$^{-1}$ to $3.50 \times 10^{-11}$ cm$^2$s$^{-1}$ (Sherratt and Murray (1991)) | $3.50 \times 10^{-11}$ cm$^2$s$^{-1}$ |
| $D_n$ | diffusivity of nutrients concentration | $1.10 \times 10^{-6}$ cm$^2$s$^{-1}$ (Casciari et al., 1998) | $1.10 \times 10^{-6}$ cm$^2$s$^{-1}$ |
| $D_o$ | diffusivity of oxygen | $1.82 \times 10^{-5}$ cm$^2$s$^{-1}$ (Mueller-Klieser and Sutherland (1984)) | $8.5 \times 10^{-5}$ cm$^2$s$^{-1}$ |
| $d_p$ | apoptosis rate of proliferative cells | $4.0 \times 10^{-10}$ s$^{-1}$ (Busini et al., 2007) | $4.0 \times 10^{-10}$ s$^{-1}$ |
| $d_q$ | apoptosis rate of quiescent cells | $4.0 \times 10^{-10}$ s$^{-1}$ (Busini et al., 2007) | $4.0 \times 10^{-10}$ s$^{-1}$ |
| $\alpha$ | tumour cell proliferation rate | $1.38 \times 10^{-6}$ s$^{-1}$ to $5.92 \times 10^{-5}$ s$^{-1}$ (Burton (1966)) | $1.50 \times 10^{-5}$ s$^{-1}$ |
| $\gamma$ | quiescent to necrosis transformation rate | $3.80 \times 10^{-6}$ s$^{-1}$ (Busini et al., 2007) | $2.60 \times 10^{-5}$ s$^{-1}$ |

We have used values for different parameters from the available data in the literature, as indicated in Table 1. For a few parameters, the values were not available. Their values are fixed through trial and error. We have done our computer experiment using spatial scale $10^{-2}$ mm. For an avascular tumour the maximum radius is 1 mm. We have considered $l$ = 2.50 mm (the distance between the tumour centre to the nearest blood vessel). We assume the diffusivity of living tumour cells (either proliferative or quiescent) is the same as the diffusivity of epithelial cells. Stokes et al., (1990) have estimated the coefficient of the independently motile endothelial cells to be in the range $2 \times 10^{-9} - 10^{-8}$ cm$^2$s$^{-1}$. However, in the case of avascular tumour growth, cells migrate due to the nutrients and oxygen present in the fluid. Sherratt and Murray (1991) have estimated the diffusivity of proliferative cells to be in the range $6.9 \times 10^{-9}$ cm$^2$s$^{-1}$ to $3.5 \times 10^{-11}$ cm$^2$s$^{-1}$ at the presence of growth molecules. We take $D_{p0} = D_{q0} = 3.5 \times 10^{-11}$ cm$^2$s$^{-1}$. Mueller-Klieser and Sutherland (1984) measured the diffusivity of oxygen ($D_{o0}$) to be $1.82 \times 10^{-5}$ cm$^2$s$^{-1}$, though for our simulation oxygen diffusivity coefficient is taken as $D_{o0} = 8.5 \times 10^{-5}$ cm$^2$s$^{-1}$ and diffusivity of nutrients concentration as $D_{n0} = 1.1 \times 10^{-6}$ cm$^2$s$^{-1}$ (Casciari et al., 1998) (Refer Table 1).

Burton (1966) suggested that the tumour proliferation rate should be in the range $1.38 \times 10^{-6}$ s$^{-1}$ to $5.92 \times 10^{-5}$ s$^{-1}$. Busini et al., (2007) suggested the rate of transformation from quiescent to necrotic to be $3.80 \times 10^{-6}$ s$^{-1}$. For this simulation, we have taken proliferation rate ($\alpha$) to be $1.27 \times 10^{-5}$ s$^{-1}$ (1.10 d$^{-1}$), and quiescent to necrotic transformation rate ($\gamma$) to be $2.60 \times 10^{-5}$ s$^{-1}$ (2.25 d$^{-1}$). Transformation rate from proliferative to quiescent ($\beta$) cell has been taken as ($1.70 \times 10^{-4}$ s$^{-1}$) through trial and error as no reference for $\beta$ is available. Estimation of the parameters like, ECM velocity ($v_e$), production rates ($\mu_n$ and $\mu_o$), and consumption rates ($w_1$, $w_2$, $w_3$ and $k_1$, $k_2$, $k_3$) are very difficult to find clinically. We use $v_e = 2.4 \times 10^{-10}$ cm s$^{-1}$ and the non-dimensional rates of $\mu_n = \mu_o = 0.8510$ as well as consumption rates of $w_1 = k_1 = 0.8450$, $w_2 = k_2 = 0.8510$, and $w_3 = k_3 = 0.6380$ under the assumption that the nutrients and oxygen consumption rates of proliferative cells are higher than those of the quiescent cells.

### 3.1.2 Simulation and Results

We solve the system of Equations (12) numerically by combining Equations (7) through (11) with the initial and boundary conditions (13, 14 respectively). We assume, $t_k = k\Delta t$, $0 \le t_k \le T$, where $k = 0, 1, 2, \ldots, n_t$ ($n_t = T/\Delta t$) and $r_i = i\Delta r$, $0 \le r_i \le l$, where $i = 0, 1, 2, \ldots, n_r$ ($n_r = l/\Delta r$). Grünwald approximated these left-handed as well as the right-handed fractional derivatives (Meerschaert and Tadjeran (2006), Yang et al., 2010). By using the discretization rule we get from Equations (9) and (10),

$$\mathcal{D}_L^\phi = \frac{\partial \phi}{\partial (r)^\phi} = \frac{1}{(\Delta r)^\phi} \sum_{\chi=0}^{i+1} g_\chi p_{i-\chi+1} \text{ and } \mathcal{D}_R^\phi = \frac{\partial \phi}{\partial (-r)^\phi} = \frac{1}{(\Delta r)^\phi} \sum_{\chi=0}^{\chi-i+1} g_\chi p_{i+\chi-1} \quad (15)$$

$$\mathcal{D}_L^\xi = \frac{\partial \xi}{\partial (r)^\xi} = \frac{1}{(\Delta r)^\xi} \sum_{\chi=0}^{i+1} g_\chi p_{i-\chi+1} \text{ and } \mathcal{D}_R^\xi = \frac{\partial \xi}{\partial (-r)^\xi} = \frac{1}{(\Delta r)^\xi} \sum_{\chi=0}^{\chi-i+1} g_\chi p_{i+\chi-1} \quad (16)$$

where, $g_\chi = (-1)^\chi \frac{\Gamma(\phi+1)}{\Gamma(\chi+1)\Gamma(\phi-\chi+1)}$ and $g_\chi = (-1)^\chi \frac{\Gamma(\xi+1)}{\Gamma(\chi+1)\Gamma(\xi-\chi+1)}$ in Equation (15) and (16) respectively. (17)

$\Gamma(.)$ is the Euler gamma function, and $\Delta r$ is the uniform size of the intervals into which the spatial axis is divided. We have included the fractional derivative on the spatial domain only and applied forward differencing method for the time domain. The step sizes are considered as $\Delta t$ = 0.004 and $\Delta r$ = 1. The model is simulated with $p_0 = 1$, $q_0 = 2.25$, $n_0 = 1.50$, $c_0 = 1$, $P_1 = 1$, $\theta = 1$, and $\rho = 1.25$.

According to (Oncologist's Note: https://notesofoncologist.com/2018/02/26/how-fast-do-tumours-grow/) avascular tumour in epithelial tissue grows up to 2 mm in diameter over 6 years. In this study, we consider, an iteration is equivalent to one day. So we iterate this process for 2200 times (2200 days ≈ 6 years) and collect the result at the duration of 200 days.

At the initial phase (up to 600 days) of the simulation, proliferative cells are mostly concentrated near the centre (Fig. 3 (a)). After that, it gradually moves in the forward direction, and at the end of the simulation (after 2200 days) it reaches 1.44 mm (approx.) from the centre (concentration value up to 0.001 considered). It indicates that after 6 years, the radius of the tumour will be 1.44 mm (approx.). Up to 600 days there is no sign of quiescent cells. After 1000 days and onwards the quiescent cells gradually increase (Fig. 3 (b)) due to steady fall of nutrients concentration (Fig. 5 (a)) and partial pressure of oxygen (Fig. 5 (b)) within the tumour cells nearer to the centre. After 1200 days, oxygen and nutrients levels further decrease sharply near the centre of the tumour. As a result, quiescent cells (hypoxic) near the centre are transformed into necrotic cells. With time the necrotic core increases rapidly and reaches approximately 0.80 mm (Fig. 4 (a)) in radius, whereas hypoxic cells grow approximately 1.10 mm from the tumour centre (Fig. 3 (b)). While the outer surface of the tumour always contains proliferative cells with higher concentrations, the volume of the necrotic core increases with the volume of the tumour. The overlapping areas in Fig. 4 (b) between different cell layers indicate that, boundaries between these layers are not distinguishable. Tumour regression cannot be seen in its life time. The above simulation is done with $\phi = 1.75$, $\xi = 0.75$, and $\psi = 0.5$.

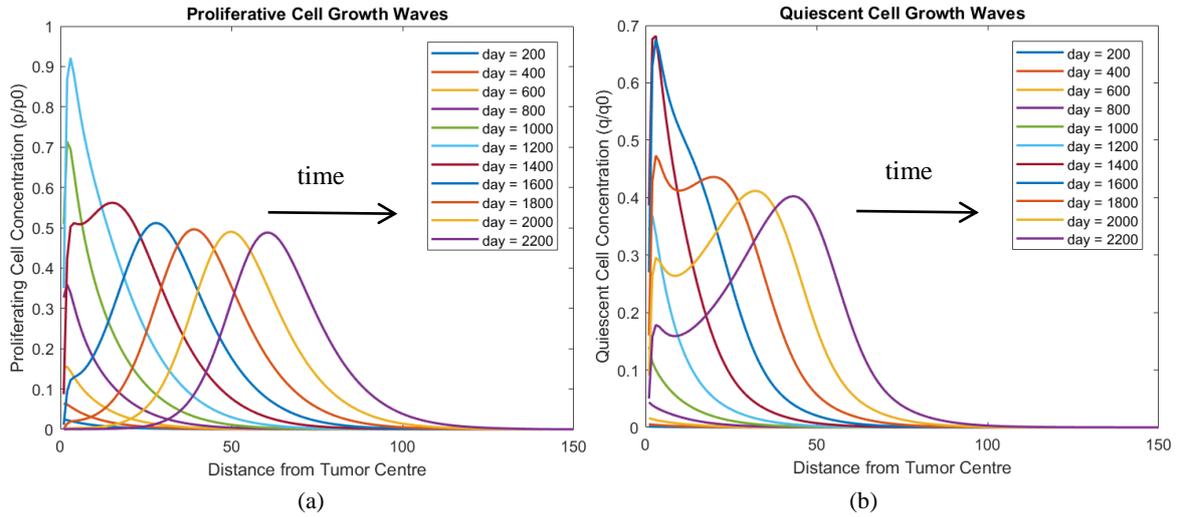

**Fig. 3** (a) Proliferative ($p/p_0$), (b) quiescent ($q/q_0$) cell concentration waves at different time intervals with respect to the distance from the tumour centre.

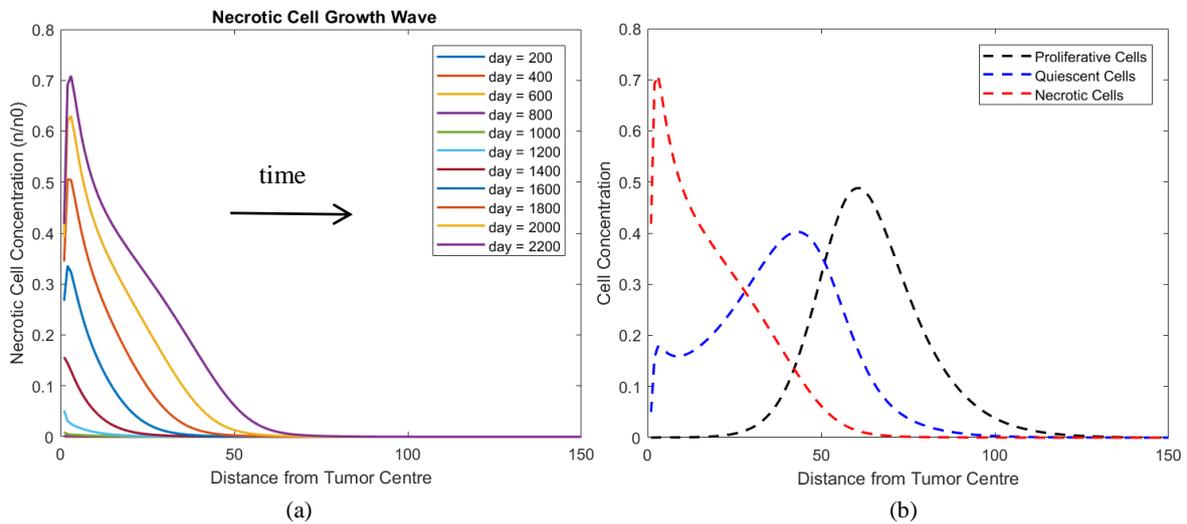

**Fig. 4** (a) Necrotic ($n/n_0$) cell concentration waves at different time intervals with respect to the distance from tumour centre, and (b) represents proliferative, quiescent, and necrotic cell concentration wave on 2200$^{th}$ day.

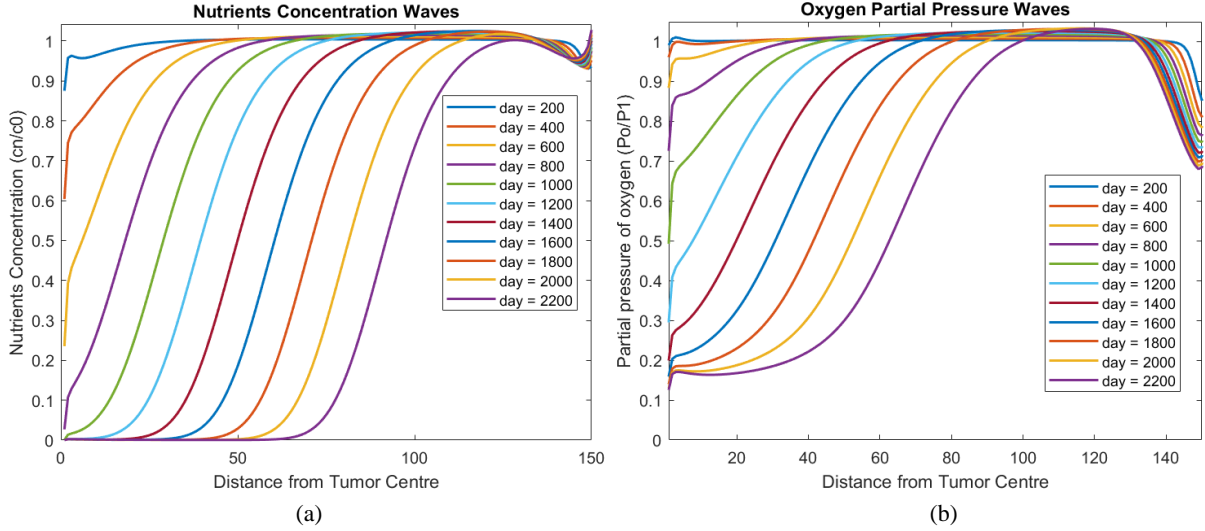

**Fig. 5** Represent (a) nutrients concentration ($c_n/c_0$) and (b) partial pressure of oxygen ($P_o/P_1$) waves at different time intervals.

### 3.2 Variable Fractional Order ADE Model

Though FO-FADE have shown some advantages over simple ADE (Caputo and Cametti (2008); Morales-Casique et al., 2006a; Morales-Casique et al., 2006b); Cushman and Ginn (2000)) to model anomalous diffusive processes, but in the case of biological phenomena, it has been suggested to use VO-FADE based model (de Azevedo et al., 2006; Atangana and Secer (2013)) as the cellular membrane is porous in nature and also the structure of the medium or external fields change with time (Atangana and Secer (2013)). We modify our proposed model (12) by including VO-FADE. In this context, we assume that $\phi = \phi(r,t)$ and also $\xi = \xi(r,t)$ in Equation (8). Now, the space fractional derivative is included in the model from (Zhuang et al. 2009), and we get,

$$\frac{\partial^{\phi(r_i,t_k)} p}{\partial r^{\phi(r_i,t_k)}} = \frac{\partial^{\phi_i^k} p}{\partial r^{\phi_i^k}} \approx \left(\left(\frac{1+\psi}{2}\right) \mathcal{D}_L^{\phi_i^k}(p) + \left(\frac{1-\psi}{2}\right) \mathcal{D}_R^{\phi_i^k}(p)\right), \; 1 < \phi(r,t) \leq 2 \quad (18)$$

and $\quad \dfrac{\partial^{\xi(r_i,t_k)} p}{\partial r^{\xi(r_i,t_k)}} = \dfrac{\partial^{\xi_i^k} p}{\partial r^{\xi_i^k}} \approx \left(\left(\dfrac{1+\psi}{2}\right) \mathcal{D}_L^{\xi_i^k}(p) - \left(\dfrac{1-\psi}{2}\right) \mathcal{D}_R^{\xi_i^k}(p)\right), \; 0 < \xi(r,t) \leq 1 \quad (19)$

where, $\phi(r_i,t_k)$ and $\xi(r_i,t_k)$ are the corresponding fractional orders at the $i^{th}$ radial point at the $k^{th}$ time instance, and $\psi \; (-1 \leq \psi \leq 1)$ is the skewness parameter. $\mathcal{D}_L^{\phi_i^k} (\mathcal{D}_L^{\xi_i^k})$ and $\mathcal{D}_R^{\phi_i^k} (\mathcal{D}_R^{\xi_i^k})$ are the left- and right-handed fractional derivative of $\phi(r_i,t_k)$ and $\xi(r_i,t_k)$ respectively. $L$ and $R$ are the corresponding lower and upper bounds of $\phi(r,t)$ ($\xi(r,t)$). The definitions of left- and right-hand limits are,

$$\mathcal{D}_L^{\phi_i^k} = \frac{\partial^{\phi_i^k}}{\partial (r)^{\phi_i^k}} \quad \text{and} \quad \mathcal{D}_R^{\phi_i^k} = \frac{\partial^{\phi_i^k}}{\partial (-r)^{\phi_i^k}} \quad (20)$$

$$\mathcal{D}_L^{\xi_i^k} = \frac{\partial^{\xi_i^k}}{\partial (r)^{\xi_i^k}} \quad \text{and} \quad \mathcal{D}_R^{\xi_i^k} = \frac{\partial^{\xi_i^k}}{\partial (-r)^{\xi_i^k}} \quad (21)$$

#### 3.2.1 Simulation and Results

We also solve this model (12) numerically by combining Equations (18) through (21) with the initial condition (13) and boundary conditions (14). We assume, $t_k = k\Delta t$, $0 \leq t_k \leq T$, where $k = 0, 1, 2, \ldots, n_t$ and $r_i = i\Delta r$, $0 \leq r_i \leq l$, where $i = 0, 1, 2, \ldots, n_r$. We discretize the Equations (20) through (21) (Zhuang et al. 2009),

$$\mathcal{D}_L^{\phi_i^k} p(r_i, t_k) = \mathcal{D}_L^{\phi_i^k} p_i^k = \frac{1}{(\Delta r)^{\phi_i^k}} \sum_{\chi=0}^{i+1} g_{\phi_i^k}^{(\chi)} p_{i-\chi+1}^k \qquad (22)$$

$$\mathcal{D}_R^{\phi_i^k} p(r_i, t_k) = \mathcal{D}_R^{\phi_i^k} p_i^k = \frac{1}{(\Delta r)^{\phi_i^k}} \sum_{\chi=0}^{n_r-i+1} g_{\phi_i^k}^{(\chi)} p_{i+\chi-1}^k \qquad (23)$$

$$\mathcal{D}_L^{\xi_i^k} p(r_i, t_k) = \mathcal{D}_L^{\xi_i^k} p_i^k = \frac{1}{(\Delta r)^{\xi_i^k}} \sum_{\chi=0}^{i+1} g_{\xi_i^k}^{(\chi)} p_{i-\chi+1}^k \qquad (24)$$

$$\text{and} \quad \mathcal{D}_R^{\xi_i^k} p(r_i, t_k) = \mathcal{D}_R^{\xi_i^k} p_i^k = \frac{1}{(\Delta r)^{\xi_i^k}} \sum_{\chi=0}^{n_r-i+1} g_{\xi_i^k}^{(\chi)} p_{i+\chi-1}^k \qquad (25)$$

$$\text{where,} \; g_{\phi_i^k}^{(\chi)} = (-1)^\chi \frac{\Gamma(\phi_i^k+1)}{\Gamma(\chi+1)\Gamma(\phi_i^k-\chi+1)}, \text{ and } g_{\xi_i^k}^{(\chi)} = (-1)^\chi \frac{\Gamma(\xi_i^k+1)}{\Gamma(\chi+1)\Gamma(\xi_i^k-\chi+1)} \qquad (26)$$

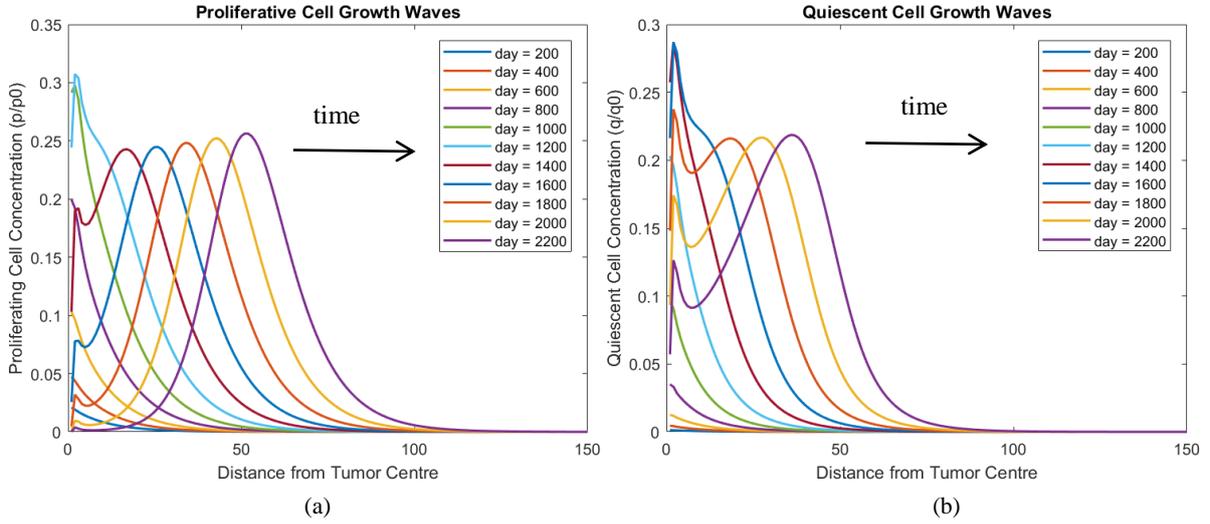

**Fig. 6** (a) Proliferative ($p/p_0$), (b) quiescent ($q/q_0$) cell concentration waves at different time intervals with respect to the distance from the tumour centre.

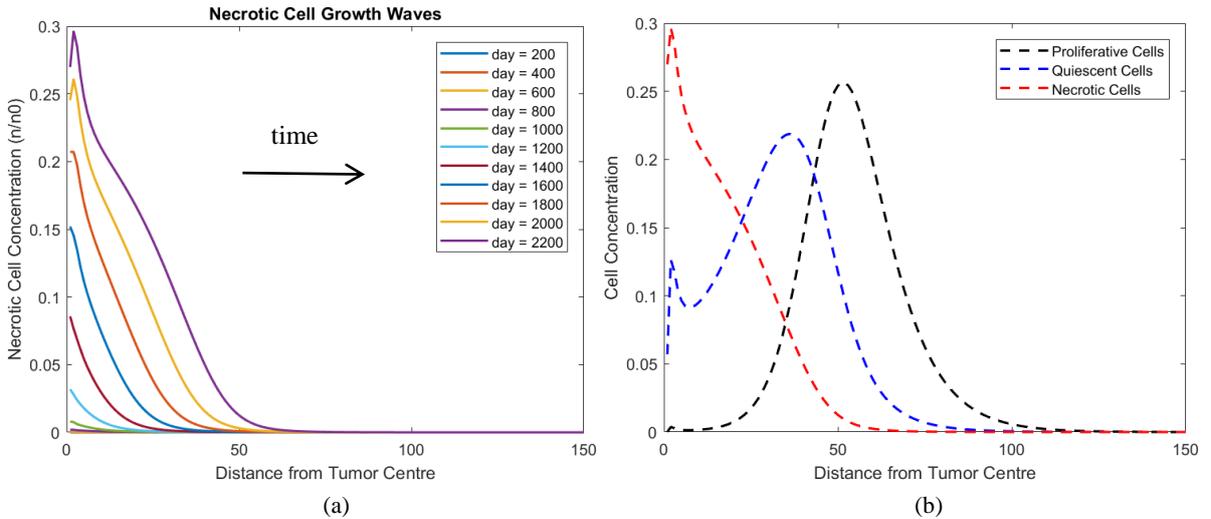

**Fig. 7** (a) Necrotic ($n/n_0$) cell concentration waves at different time intervals with respect to the distance from tumour centre, and (b) represents proliferative, quiescent, and necrotic cell concentration wave on 2200$^{th}$ day.

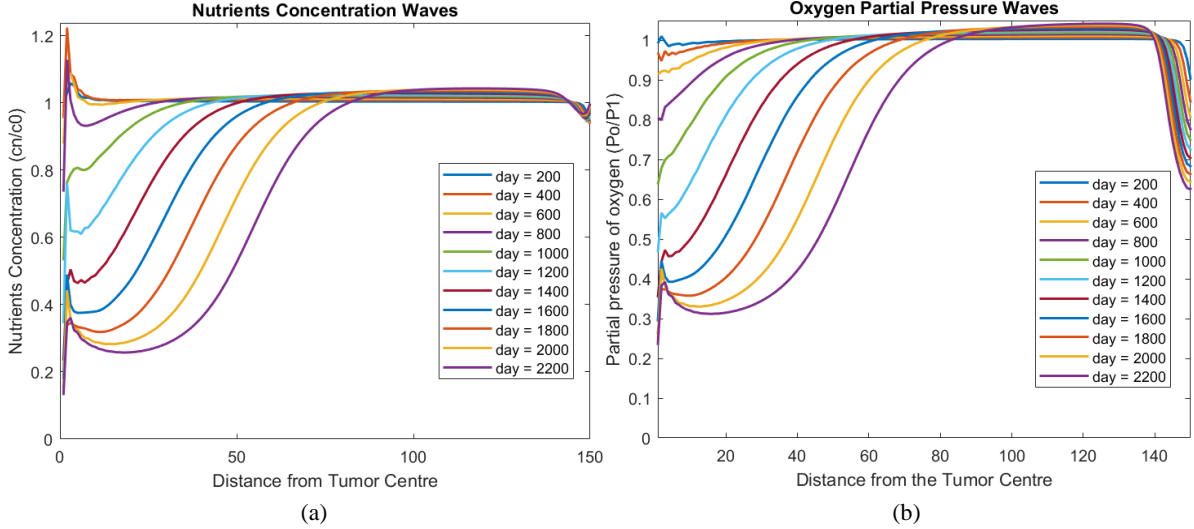

**Fig. 8** Represent (a) nutrients concentration ($c_n/c_0$) and (b) partial pressure of oxygen ($P_o/P_1$) waves at different time intervals.

We restrict the value of $\phi_i^k$ ($1.55 \leq \phi_i^k \leq 1.95$) and $\xi_i^k$ ($0.55 \leq \xi_i^k \leq 0.95$) for the simulation. Therefore, we assume, $\phi_i^k = 1.75 + 0.2 * \sin(0.5\pi r_i t_k)$, and $\xi_i^k = 0.75 + 0.2 * \sin(0.5\pi r_i t_k)$. We have applied the fractional derivative on the spatial domain only. The previously used parameter values (mentioned in sec. 3.1.1) are used for the simulation. We have iterated the simulation process for 2200 times and collect the result at the interval of 200 days.

Fig. 6 (a) describes that from 600 days onwards the proliferative cells wave grows up and within six years it reaches up to 1.18 mm (approx.). On the other hand, quiescent cells gradually increase after 1000 days (Fig. 6 (b)) due to the steady fall of nutrients concentration (Fig. 8 (a)) and partial pressure of oxygen (Fig. 8 (b)) within the tumour cells nearer to the centre. After 1400 days, oxygen and nutrients levels decrease heavily among the tumour cells; as a result quiescent cells nearer to the tumour centre are transformed into necrotic cells in a rapid speed. With time the necrotic core reaches approximately 0.75 mm (Fig. 7 (a)) in radius, whereas quiescent cells grow approximately 1.0 mm from the tumour centre (Fig. 6 (b)). The outer surface of the tumour always contains proliferative cells with higher concentration. The overlapping areas in Fig 7 (b) indicates that boundary among these layers are not distinguishable. Tumour regression cannot be seen in its life time. All the simulations have been carried out in MATLAB R2017a on a Pentium $i$5 processor. The pseudo-code for both the models is given in the Appendix.

## 3.3 Discussion

We also compare both of these models from phenomenological point of view. The tumour radius under the FO-FADE model is 1.44 mm. and that under the VO-FADE model is 1.18 mm. We also determine the tumour volume at different time intervals and the volume growth is shown in Fig. 9 (a) for both of these models. It is observed that at the initial phase, the rate of increment in tumour volume in the FO-FADE model is higher than that in the VO-FADE model. After 1800 days, the simulated volume becomes stagnant in the FO-FADE model, but there is no change in the rate of increment of volume under the VO-FADE model. We also determine the total number of tumour cells (= tumour volume × avg. cell concentrations) and plotted in Fig. 9 (b) at different time intervals. It is observed that up to 600 days the number of cells in both of the models are very low, after that they grow at a higher rate in the FO-FADE model than that in the VO-FADE model. At the end of simulation (2200 days), the number of cells under the FO-FADE model is $3.5 \times 10^6$, whereas that in the VO-FADE model is $1.13 \times 10^6$. At the macroscopic level, both of these models approximate the physical phenomena well and in consonance with the clinical facts. Though, the FO-FADE model overestimates the computed parameters (tumour radius, tumour volume, and total tumour cells) than the VO-FADE model does.

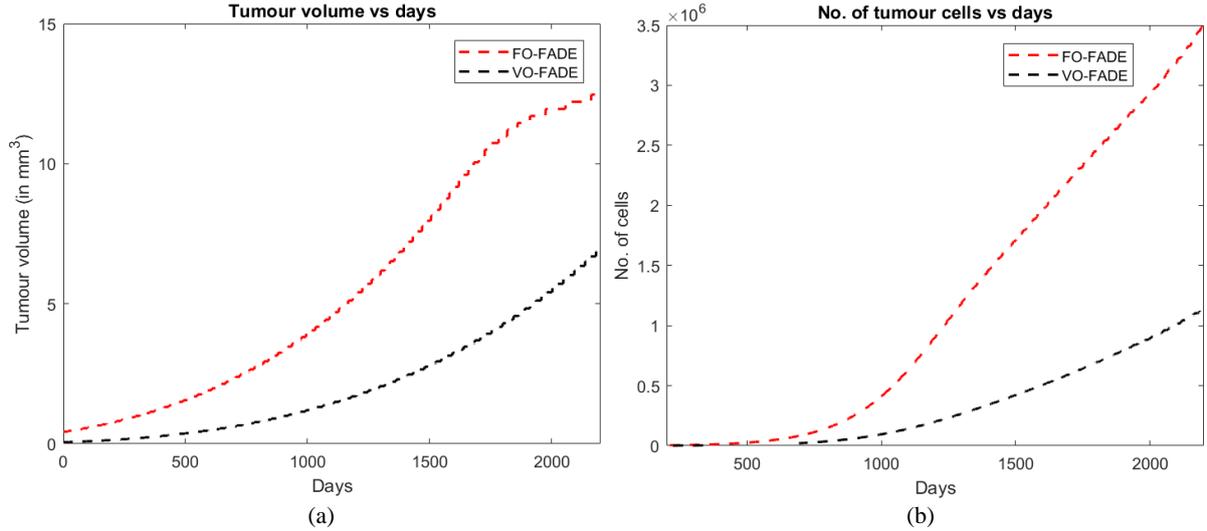

**Fig. 9** Represent (a) tumour volume (in mm³), (b) total tumour cells in different time intervals for FO-FADE and VO-FADE based models.

We measure VDT for both of these models by varying tumour proliferation rate ($\alpha = 0.6$, through 1.6 at a step of 0.1) and find that the tumour VDT varying inversely with the proliferation rate. In the FO-FADE model VDT is always lower than in the VO-FADE model and the average VDT for FO-FADE is 214 days (approx.) and that in VO-FADE is 241 days (approx.), in agreement with the clinical findings. Table 2 summarizes our findings and compares clinical observations reported in the literature.

We study the behaviour of FO-FADE model by changing its order ($\phi = 1.55$, through 1.95 at a step of 0.1 by keeping skewness $\psi = 0.5$ fixed) and also by changing the skewness parameter ($\psi$ is 0.5, 0.95, –0.95 and by keeping $\phi = 1.75$ fixed). In all the occasions, we find that there are no significant changes in the radial direction, but with the increment of the order ($\phi$) in the FO-FADE model only concentration levels (proliferative cells, quiescent cells, and necrotic cells) increase and no significant changes are visible for changing the skewness parameter $\psi$.

For model validation, one of the popular schemes is sensitivity analysis (Basu and Roy (2004)). It is used to study the robustness of a model by changing one or more of the input parameters. If the outcome of the model changes significantly with the input parameter values (Law et al., 2000; Storer et al., 2003; Ehrlén et al., 2001; Jenerette and Wu (2001)), the model is strongly sensitive to the parameter values indicating scope for refinement of the model's underlying assumptions. There are various types of sensitivity analysis methods available like, mathematical, statistical, and graphical methods. These methods are different in term of applicability, computational issues, complexity of their applications, and representations.

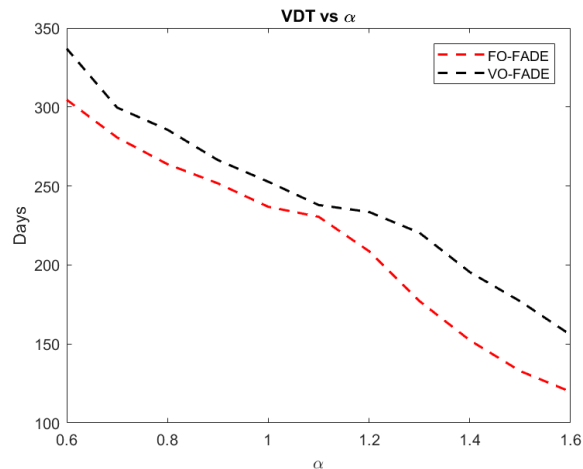

**Fig. 10** Represents tumour volume doubling time (VDT) with respect to proliferation rate ($\alpha$) in FO-FADE and VO-FADE based models.

Table 2 Tumour variables outcome in the FO-FADE and the VO-FADE model

| Tumour variable | FO-FADE | VO-FADE | Clinical observation | |
|---|---|---|---|---|
| Radius | 1.44 mm | 1.18 mm | Up to 1 mm | (Zetter (1998); Hillen and Griffioen(2007)) |
| Volume | 12.47 mm$^3$ | 6.86 mm$^3$ | 4.17 mm$^3$ | (Zetter (1998); Hillen and Griffioen(2007)) |
| No. of tumour cells | 3.50×10$^6$ | 1.13×10$^6$ | 10$^6$ to 10$^7$ | (Lodish et al., 2000; Anderson (2005)) |
| VDT | 214 days | 241 days | 150 to 400 days (Kanashiki et al., 2012) | |

Table 3 Measuring RMSD for FO-FADE and VO-FADE model by changing variables

| Model | Tumour parameters | 5% increment | 10% increment | Without changing parameters | 5% decrement | 10% decrement | RMSD | RMSD (in %) |
|---|---|---|---|---|---|---|---|---|
| FO-FADE | Radius (mm) | 1.46 | 1.46 | 1.44 | 1.43 | 1.41 | 0.0210 | 1.46 |
| VO-FADE |  | 1.21 | 1.28 | 1.18 | 1.13 | 1.08 | 0.0760 | 6.46 |
| FO-FADE | Volume (mm$^3$) | 12.99 | 12.99 | 12.47 | 12.21 | 11.70 | 0.54 | 4.32 |
| VO-FADE |  | 7.39 | 8.75 | 6.86 | 6.02 | 5.26 | 1.33 | 19.39 |
| FO-FADE | Total cells | 3.96x10$^6$ | 3.81x10$^6$ | 3.50x10$^6$ | 3.19x10$^6$ | 2.90x10$^6$ | 3.19 x10$^5$ | 9.18 |
| VO-FADE |  | 1.28x10$^6$ | 1.29x10$^6$ | 1.13x10$^6$ | 9.93x10$^5$ | 8.82x10$^5$ | 1.79 x10$^5$ | 16.05 |
| FO-FADE | Avg. VDT (in days) | 218 | 198 | 214 | 234 | 243 | 19.44 | 8.78 |
| VO-FADE |  | 239 | 238 | 241 | 250 | 258 | 9.78 | 3.98 |

Here, we make a partial sensitivity analysis (Table 3) by calculating the effect of the model outcome through changing some input parameters, while other parameters are kept fixed. The sensitiveness of both the models are represented by the root mean squared deviation (RMSD) percentage with respect to the mean output value of the model. The analysis is repeated for a couple of times by changing the values of the parameters $α$, $β$, $γ$, $μ_n$, $μ_o$, $w_1$, $w_2$, $w_3$, $k_1$, $k_2$, and $k_3$ by changing them $±5\%$, $±10\%$ by keeping the other parameter values fixed. The growth waves for proliferative, quiescent, and necrotic cells are neither distorted nor any significant changes are visible in their patterns after changing the values of the parameters $α$, $β$, $γ$, $μ_n$, $μ_o$, $w_1$, $w_2$, $w_3$, $k_1$, $k_2$, and $k_3$ by $±5\%$, $±10\%$. In the FO-FADE model, the RMSD percentage for radius is 1.46, for volume 4.32, total number of tumour cells 9.18, and average VDT 8.78. On the other hand, in the VO-FADE model the RMSD percentage for radius is 6.46, volume 19.39, total number of tumour cells 16.05, and average VDT 3.98. The RMSD percentages indicate that the sensitivity of FO-FADE is very low, whereas VO-FADE model is moderately sensitive to the parameter values.

## 4  Conclusions

Over the decades there has been growing interest in using mathematical modeling for refining our understanding of many clinical problems. Diffusion is an important process in these models. Diffusion processes in biological systems are heterogeneous in nature as the cellular membranes are porous. Scientists have advocated for using FADE based model over simple ADE or DE-based model to explain heterogeneous biological systems. Though, most of the researchers neglect this fact and use simple ADE or simple DE to model tumour growth process.

In this paper, we have developed two models: FO-FADE based model and VO-FADE based model for avascular tumour growth. We study both the models from phenomenological and clinical point of views by measuring tumour radius, tumour volume, and total number of cells in the tumour over time and also determine tumour VDT. We find that both of these models offers realistic and insightful information for tumour growth at the macroscopic level, and well approximate the physical phenomena. FO-FADE based model always overestimates clinical facts like tumour radius, tumour volume and total cell counts, whereas VO-FADE based model always justify the clinical facts in the tumour growth phenomena. As the simulation parameters may change due to different biochemical and biophysical processes, the robustness of the model is determined by changing the values of the important parameters. It is found that the FO-FADE model is more robust than the VO-FADE model.

**Appendix**

**Pseudo-code (VO-FADE model)**

Step 1: Initialize all the variables $D_p$, $D_q$, $D_n$, $D_o$, $\alpha$, $\beta$, $\gamma$, $v_e$, $\mu_n$, $\mu_o$, $w_1$, $k_1$, $w_2$, $k_2$, $w_3$, $k_3$, $l$, $\Delta r$, $T$, $\Delta t$, $p_0$, $q_0$, $n_0$, $c_0$, $P_1$, $\theta$, $d_p$, $d_q$, $n_r$, and $n_t$, $r[n_r] = [0, 1, 2, …, l]$, $t[n_r] = [\Delta t, 2\Delta t, …, T]$;    // $n_r = l/\Delta r$ and $n_t = T/\Delta t$

Step 2: Initialize $\rho$, and $\psi$. And also initialize $\phi[n_t][n_r]$, and $\xi[n_t][n_r]$ using appropriate functions;

Step 2: Transform all the variables in a same scale;    //distance in $10^{-2}$ mm, and time in second

Step 3: Non-dimensionalize the variables;    //for the clarity we drop the * from the variables

Step 4: Initialize $p$, $q$, $n$, $c_n$, and $P_o$;

Step 5: Set boundary values of $p$, $q$, $n$, $c_n$, and $P_o$;

Step 5: Create arrays: $p[n_r]$, $q[n_r]$, $n[n_r]$, $c_n[n_r]$, and $P_o[n_r]$;
      Temporary arrays: $nextp[n_r]$, $nextq[n_r]$, $nextn[n_r]$, $nextc_n[n_r]$, and $nextP_o[n_r]$;
      Auxiliary arrays: $Ap[n_t][n_r]$, $Aq[n_t][n_r]$, $An[n_t][n_r]$, $Ac_n[n_t][n_r]$, and $AP_o[n_t][n_r]$;

Step 6: Set the boundary values for $p$, $q$, $n$, $C_n$ and $P_o$;

Step7: *For $k = 1$ to $n_t$*
    *For $i = 1$ to $n_r-2$*
        $nextp[i] = p[i]+ \Delta t$ ($A+ \alpha h_1(c_n, P_o) p[i]- \beta h_2(c_n, P_o) p[i]-d_p p[i]$);
        (Like the same way determine $nextq[n_r]$, $nextn[n_r]$, $nextc_n[n_r]$, and $nextP_o[n_r]$).
    *Endfor*
    $p[1 : n_r-2] = nextp[1 : n_r-2]$;
    $Ap[k][0:n_r-1] = p[0 : n_r-1]$;
    (Same way update $q$, $Aq$, $n$, $An$, $c_n$, $Ac_n$, $P_o$, and $AP_o$ respectively).
*Endfor*

Step 8: End

In Step7, $A = D_p h^\rho (c_n, P_o)(t_1+(2/r[i])t_2) -v_e h(c_n, P_o)(t_2+(2/r[i])p[i])-D_p(\rho c_n [i] h^\rho (c_n, P_o)t_3+\rho P_o[i] h^\rho (c_n, P_o)t_4)t_2$
$-v_e(\rho c_n[i] h^\rho (c_n, P_o)t_3-\rho P_o[i] h^\rho (c_n, P_o)t_4)p[i]$;

where, $t_1= 0.5(1+ \psi)(1/(\Delta r^{\phi[k][i]}))H_1+ 0.5(1+ \psi)(1/(\Delta r^{\phi[k][i]}))H_2$;    // compute $\frac{\partial^{\phi[k][i]} p}{\partial r^{\phi[k][i]}}$

$t_2 = 0.5(1+ \psi)(1/(\Delta r^{\xi[k][i]}))H_3- 0.5(1+ \psi)(1/(\Delta r^{\xi[k][i]}))H_4$;    // compute $\frac{\partial^{\xi[k][i]} p}{\partial r^{\xi[k][i]}}$

$t_3 = 0.5(1+ \psi)(1/(\Delta r^{\xi[k][i]}))H_5- 0.5(1+ \psi)(1/(\Delta r^{\xi[k][i]}))H_6$;    // compute $\frac{\partial^{\xi[k][i]} P_o}{\partial r^{\xi[k][i]}}$

$t_4 = 0.5(1+ \psi)(1/(\Delta r^{\xi[k][i]}))H_7- 0.5(1+ \psi)(1/(\Delta r^{\xi[k][i]}))H_8$;    // compute $\frac{\partial^{\xi[k][i]} c_n}{\partial r^{\xi[k][i]}}$

$H_1$, $H_2$, …, $H_8$ are determined following way:

*for $j = 0$ to $i+1$*
    $H_1 = H_1 + (-1)^j \frac{\Gamma(\phi[k][i]+1)}{\Gamma(j+1)\Gamma(\phi[k][i]+1-j)} p[i-j+1]$;    // compute $\mathcal{D}_L^{\phi[k][i]}(p) = \frac{\partial^{\phi[k][i]} p}{\partial (r)^{\phi[k][i]}} = \frac{1}{(\Delta r)^{\phi[k][i]}} \sum_{\chi=0}^{i+1} g_\chi p_{i-\chi+1}$
*Endfor*

*for $j = 0$ to $n_r-i+1$*

$H_2 = H_2 + (-1)^j \frac{\Gamma(\phi[k][i]+1)}{\Gamma(j+1)\Gamma(\phi[k][i]+1-j)} p[i+j-1];$ // compute $\mathcal{D}_R^{\phi[k][i]}(p) = \frac{\partial^{\phi[k][i]} p}{\partial (-r)^{\phi[k][i]}} = \frac{1}{(\Delta r)^{\phi[k][i]}} \sum_{\chi=0}^{\chi-i+1} g_\chi p_{i+\chi-1}$

*Endfor*

*for j = 0 to i+1*

$H_3 = H_3 + (-1)^j \frac{\Gamma(\xi[k][i]+1)}{\Gamma(j+1)\Gamma(\xi[k][i]+1-j)} p[i-j+1];$ // compute $\mathcal{D}_L^{\xi[k][i]}(p) = \frac{\partial^{\xi[k][i]} p}{\partial (r)^{\xi[k][i]}} = \frac{1}{(\Delta r)^{\xi[k][i]}} \sum_{\chi=0}^{i+1} g_\chi p_{i-\chi+1}$

*Endfor*

*for j = 0 to $n_r$–i+1*

$H_4 = H_4 + (-1)^j \frac{\Gamma(\xi[k][i]+1)}{\Gamma(j+1)\Gamma(\xi[k][i]+1-j)} p[i+j-1];$ // compute $\mathcal{D}_R^{\xi[k][i]}(p) = \frac{\partial^{\xi[k][i]}}{\partial (-r)^{\xi[k][i]}} = \frac{1}{(\Delta r)^{\xi[k][i]}} \sum_{\chi=0}^{\chi-i+1} g_\chi p_{i+\chi-1}$

*Endfor*

*for j = 0 to i+1*

$H_5 = H_5 + (-1)^j \frac{\Gamma(\xi[k][i]+1)}{\Gamma(j+1)\Gamma(\xi[k][i]+1-j)} P_o[i-j+1];$ // compute $\mathcal{D}_L^{\xi[k][i]}(P_o) = \frac{\partial^{\xi[k][i]} P_o}{\partial (r)^{\xi[k][i]}} = \frac{1}{(\Delta r)^{\xi[k][i]}} \sum_{\chi=0}^{i+1} g_\chi P_{o_{(i-\chi+1)}}$

*Endfor*

*for j = 0 to $n_r$–i+1*

$H_6 = H_6 + (-1)^j \frac{\Gamma(\xi[k][i]+1)}{\Gamma(j+1)\Gamma(\xi[k][i]+1-j)} P_o[i+j-1];$ //compute $\mathcal{D}_R^{\xi[k][i]}(P_o) = \frac{\partial^{\xi[k][i]} P_o}{\partial (-r)^{\xi[k][i]}} = \frac{1}{(\Delta r)^{\xi[k][i]}} \sum_{\chi=0}^{\chi-i+1} g_\chi P_{o_{(i+\chi-1)}}$

*Endfor*

*for j = 0 to i+1*

$H_7 = H_7 + (-1)^j \frac{\Gamma(\xi[k][i]+1)}{\Gamma(j+1)\Gamma(\xi[k][i]+1-j)} c_n[i-j+1];$ //compute $\mathcal{D}_L^{\xi[k][i]}(c_n) = \frac{\partial^{\xi[k][i]} c_n}{\partial (r)^{\xi[k][i]}} = \frac{1}{(\Delta r)^{\xi[k][i]}} \sum_{\chi=0}^{i+1} g_\chi c_{n_{(i-\chi+1)}}$

*Endfor*

*for j = 0 to $n_r$–i+1*

$H_8 = H_8 + (-1)^j \frac{\Gamma(\xi[k][i]+1)}{\Gamma(j+1)\Gamma(\xi[k][i]+1-j)} c_n[i+j-1];$ //compute $\mathcal{D}_R^{\xi[k][i]}(c_n) = \frac{\partial^{\xi[k][i]} c_n}{\partial (-r)^{\xi[k][i]}} = \frac{1}{(\Delta r)^{\xi[k][i]}} \sum_{\chi=0}^{\chi-i+1} g_\chi c_{n_{(i+\chi-1)}}$

*Endfor*

FO-FADE model is the simplified version of VO-FADE model. The only difference between them is that, in VO-FADE the order of the derivatives changes with the iterative variable *k* and *i*, but in FO-FADE model the order is fixed for the simulation.

**Acknowledgement**

We are sincerely thankful to Dr. Naveen Kumar, Department of Mathematics, Banaras Hindu University, Varanasi 221005, India for his cooperation and help. The first author is also thankful to the University Grants Commission, Government of India, for supporting him by a Junior Research Fellowship.